\documentstyle[11pt,aaspp4]{article}

\lefthead{Cunha et al.}
\righthead{Copper in $\omega$~Centauri}

\begin{document}

\title{The Evolution of Copper in the Globular
Cluster $\omega$ Centauri}

\author{Katia Cunha}
\affil{Observat\'orio Nacional, Rua General Jos\'e Cristino 77, 20921-400,
S\~ao Crit\'ov\~ao, RJ, Brazil; katia@on.br}

\author{Verne V. Smith}
\affil{Department of Physics, University of Texas El Paso, El Paso, TX 79968 
USA; verne@barium.physics.utep.edu}

\author{Nicholas B. Suntzeff}
\affil{Cerro Tololo Interamerican Observatory, Casilla 603, La Serena, Chile;
nsuntzeff@noao.edu}

\author{John E. Norris}
\affil{Research School of Astronomy \& Astrophysics, The Australian
National University, Mt. Stromlo Observatory, Cotter Road, Weston, 
ACT 2611, Australia; jen@mso.anu.edu.au}

\author{Gary S. Da Costa} 
\affil{Research School of Astronomy \& Astrophysics, The Australian
National University, Mt. Stromlo Observatory, Cotter Road, Weston,
ACT 2611, Australia; gdc@mso.anu.edu.au}

\author{Bertrand Plez}
\affil{GRAAL, Universite Montpellier II, cc072, 34095 Montpellier cedex 05,
France; plez@graal.univ-montp2.fr}

\begin{abstract}
Copper abundances are presented for 40 red-giant
members of the massive Galactic globular cluster $\omega$~Centauri,  
as well as 15 red-giant members of the globular clusters NGC288, NGC362, 
NGC3201, NGC6752, and M4 (NGC6121).
The spectra are of relatively high spectral-resolution and
signal-to-noise.
Using these abundances, plus published literature values for field stars,
the abundance trends of [Cu/Fe] are defined as a function of [Fe/H]. 
The lowest metallicity stars in $\omega$ Cen 
have [Fe/H] $\sim$ -2.0, with the 
stars in this sample spanning a range from 
[Fe/H] $\sim$ -2.0 to -0.8.  
In the field star sample, [Cu/Fe] rises from about -0.8, at [Fe/H]= -2.5,
to about -0.4 at [Fe/H]$\sim$ -1.4, and then rises rapidly to  
[Cu/Fe]$\sim$0.0 at [Fe/H]=-1.1.  The globular clusters (other than
$\omega$ Cen) tend to also 
follow the trend as displayed by the field stars.  Unlike the field stars, 
however, $\omega$ Cen displays a constant ratio of 
[Cu/Fe]$\sim$ -0.5 all the way to [Fe/H]=-0.8.  At the metallicity of
[Fe/H]= -0.8, the values of [Cu/Fe] in $\omega$ Cen fall below the
corresponding mean ratio in the field stars by roughly 0.5 dex.  
If copper is produced primarily in type Ia supernovae, as suggested
in the literature, the lack
of an increase in [Cu/Fe] in $\omega$ Cen would suggest very little
contribution from SN Ia to its chemical evolution within the metallicity
range from [Fe/H] of -2.0 up to -0.8. 
\end{abstract}

\keywords{globular clusters: individual ($\omega$~Centauri; NGC288; NGC362;
NGC3201; NGC6752; M4)} 

\clearpage

\section{Introduction}

Omega Centauri is the most massive globular cluster known in the Milky Way.
This cluster is
also the only one known whose stars display a large range of 
abundances in all elements, such as Fe, Ca, Ti, Ni or O.  The
abundance of a standard metallicity indicator, such as Fe, spans a range
from [Fe/H]$\sim$  -2.0 up to -0.4 (Pancino et al. 2000).
The abundance variations arise from chemical self-enrichment and 
evolution in $\omega$ Cen, with many of these elements produced in supernovae
of type II (SN II), e.g. as discussed by Brown et al. (1991) or
Brown \& Wallerstein (1993).
In addition to self-enrichment in elements produced in SN II, $\omega$
Cen displays an even larger abundance range in the neutron-capture s-process
elements, such as Y, Zr, Ba, or La.  Ratios, such as [Ba/Fe], increase
tremendously as [Fe/H] increases in $\omega$ Cen: this cluster 
underwent an enormous increase in s-process abundances during its 
chemical evolution, with the s-process elements being produced in
asymptotic giant branch (AGB) stars (Lloyd Evans 1983).

Other elements, produced in sites other than SN II or AGB stars, can
provide additional insights into the chemical evolution and possible
progenitor identity of this relatively
small stellar system.  In this study, we report on copper abundances 
in 40 red giant stars in $\omega$ Cen, (10 of which have already been
studied in Smith et al. 2000), as well as in
small samples of red giants in the other globular clusters NGC288, NGC362,
NGC3201, NGC6752, and M4 (NGC6121).  
Copper abundances in these various
globular cluster giants are then compared to samples of field stars from 
the literature, in order to explore the behavior of Cu with metallicity
in the Galaxy.
The [Cu/Fe] ratio is not constant in halo stars, as shown by Sneden \&
Crocker (1988) and Sneden et al. (1991),
where [Cu/Fe] increases with increasing metallicity
as [Cu/Fe] $\alpha$ 0.4[Fe/H] up to [Fe/H]=-1, where it then reaches the
solar ratio. For higher metallicities, the [Cu/Fe] is solar. These
trends were seen both in field and globular cluster stars.

Given the somewhat complicated chemical history of copper, both
Sneden et al. (1991) and Matteucci et al. (1993) reviewed the
nucleosynthetic sites of copper (as well as zinc). Sneden et al.
suggested that much of the copper in metal-poor stars was formed in
the {\it weak} component of the s-process, whereby a neutron capture
on Fe-peak elements is produced during the late stages of core
He-burning (Couch et al. 1974, Raiteri, et al. 1991) in massive
stars. Some copper was also expected to be formed in explosive
nucleosynthesis from SN II. At higher metallicities, they
expect that most of the copper was formed in Type Ia supernovae (SN Ia). Only
a small fraction ( $<$ 30\%) of the copper abundance in the sun can
come from s-process without gross inconsistencies with other elemental
and isotopic abundances (Raiteri et al. 1992). Because the weak
component of the s-process is a secondary mechanism for
nucleosynthesis, this site for the formation of copper agrees
qualitatively with a relationship of increasing [Cu/Fe] as a function
of [Fe/H] in metal-poor stars. 

Matteucci et al. (1993) revised the Sneden et al. (1991) model to force a
quantitative fit to the data. As they note, any model is hampered by
the poorly understood nucleosynthetic yields of Cu in SN II and SN Ia.
The conclusion of their paper is
that adequate fits to these elements in metal-poor stars could be made
only if there was significant nucleosynthesis from SN Ia
in the metallicity range -2 $<$ [Fe/H] $<$ -1. 
One theoretical prediction that
indirectly pertains to the production of copper via SN Ia is the study by
Kobayashi et al. (1998) who have revisited the models for the
formation of Type Ia supernovae. In the models of single degenerate
progenitors, the wind from the accreting white dwarf plays an important
role in regulating the slow increase in mass of the white dwarf. The
existence of this wind allows a much larger parameter space of single
degenerate progenitors to produce Type Ia supernova explosions. Their
prediction is that SN Ia will not form for $[Fe/H]<-1$,
apparently contradicting the models of Matteucci et al. (1993).

The origins of the metallicity variations and the details of
chemical evolution in $\omega$ Cen are not understood.
However, it clearly has a different chemical evolution history
from the Galactic halo and disk. Moreover, 
little observational work on copper has been published since 
Sneden et al. (1991). 
With this in mind, we set out to explore
the copper abundances in $\omega$ Cen with its different chemical history to
gain insight into the sites of copper nucleosynthesis.

\section{Observations}

The echelle spectra assembled and analyzed in this paper were obtained 
at two different 
telescopes: the Blanco 4m at CTIO and the 3.9m Anglo-Australian 
Telescope (AAT). The CTIO spectra were obtained using the cassegrain 
echelle spectrograph and
the Red Long camera (590 mm focal length).  Typical resolutions were
35,000 with wavelength coverage from about 5235 $-$ 8518\AA.
In this project, 10 
red-giant members of $\omega$ Cen were observed, along with 4 members 
each in NGC288 and M4, 3 members in NGC362, and 2 members each
in NGC3201 and NGC6752. The $\omega$ Cen giant spectra were analyzed in 
Smith et al. (2000), while the other cluster members are analyzed for 
the first time in this paper.  
All red giants observed have 
effective temperatures in the  range of roughly 3800--4500K.
The CTIO data were reduced to wavelength-calibrated spectra with
the IRAF programs. 
More details in the
data reduction procedures are described in Smith et al. (2000).  Final
signal-to-noise ratios in these spectra were typically 100 to 250.

In addition to the CTIO spectra, high-resolution (R$\sim$ 38,000)
spectra were analyzed for Cu I from the spectra previously studied
by Norris
\& Da Costa (1995) and Norris, Da Costa, \& Tingay (1995).  These two papers
provided equivalent widths and an abundance analysis of several elements
for a sample of 40 $\omega$
Cen giants, but the authors did not determine Cu abundances.  These spectra were
obtained with the University College London Echelle Spectrograph
(UCLES), with a CCD detector at the coude focus of the 3.9m Anglo
Australian Telescope (AAT).  
The data were reduced to 1-d spectra
using both IRAF and FIGARO routines: more detailed discussion of the
data reduction of the UCLES spectra can be found in Norris et al. (1995).
Typical signal-to-noise ratios of these final, reduced spectra are
S/N$\sim$ 50.  

Examples of spectra showing the Cu I line, as well as their synthetic
fits, are discussed and shown in Section 3 (Analysis).  Table 1 is a
summary of the observing log for the spectra analyzed
in this study. 

\section{Analysis}

The techniques for model atmosphere abundance analyses of red giants
spanning the effective temperature range sampled in this study are well
established and more detailed discussions of the techniques adopted here 
can be found in earlier papers: e.g., Norris \& Da Costa (1995),
Smith et al. (1995), or Smith et al. (2000). 
In short, published model atmospheres, based upon the
Bell et al. (1976: here BEGN) 
grid for red giants, are used with basic
stellar parameters ($T_{\rm eff}$, log $g$, and metallicity) derived
from a combination of photometry and spectroscopic analyses.

The abundance analysis was carried out for the program
stars under the assumption of LTE, using the most recent version of the
spectral analysis code MOOG (Sneden 1973).  Model atmospheres were
interpolated within the standard grid of BEGN models produced by
Gustafsson et al. (1975).  For red giants spanning the range of
temperatures, gravities, and metallicities that represent the
program stars, use of these models, with appropriately selected
samples of spectral lines, can produce abundance results that are
accurate (on an absolute scale) to within typically 0.1 $-$ 0.2 dex,
depending upon the spectral lines and species in question.  

The copper abundances derived in this study are based on the single Cu I line 
at 5782.127\AA, and adopting the gf-value used by Sneden \& Crocker (1988)
(same as in Sneden, Gratton, \& Crocker 1991), which itself is from 
the laboratory  
determination by Hannaford \& Lowe (1983).  In addition, hyperfine (hfs)
and isotopic splitting is significant for this line and these
components were included in the analysis, with the hfs and isotopic
data taken from Biehl (1976).   The various hyperfine and isotopic
components of the 5782\AA\ line, along with their gf-values, are listed
in Table 2: the gf-values are weighted by their respective solar system
isotopic fractions of 0.69 for $^{63}$Cu and 0.31 for $^{65}$Cu.  

A synthetic fit of the Cu I line in the Solar Flux Atlas 
(Kurucz et al. 1984) is shown in Figure 1.
The overall fit of the synthetic hfs profile with the observed solar
line-profile is quite good, although there are some small mismatches
near the line center (these will not affect measurably the derived stellar
Cu abundances). 
The resulting solar Cu abundance is log $\epsilon$ $=$ 4.06, compared 
to the accepted photospheric abundance of 4.21 $\pm$ 0.04 
(Grevesse \& Sauval 1998).  It should be noted  
that the solar Cu abundance quoted by Grevesse \& Sauval was 
derived using the gf-value from Koch \& Richter (1968), with
log (gf)$_{5782}$= -1.78 compared to -1.72 from Hannaford \& Lowe (1983).
This suggests that a comparison of the Grevesse Cu photospheric abundance 
to the one derived here should include a --0.06 dex offset from the differing
gf-values.  Also, Grevesse quoted a result based upon the empirical
Holweger--M\"uller model (Holweger \& M\"uller 1974), while we used a
theoretical MARCS model: the small remaining difference of 0.09 dex is,
thus, entirely reasonable.  Of importance for later discussion comparing
results derived here with the field-star abundances from Sneden \&
Crocker (1988) and Sneden et al. (1991), is that these studies and our
study used the same gf-value for the 5782\AA\ Cu I line.  In their analysis,
Sneden \& Crocker (1988) derived a solar photospheric copper abundance
using the same gf-value as adopted here, and found log $\epsilon$= 4.12.
They used the Holweger--M\"uller solar model atmosphere and the 
small difference of 0.06 dex between these respective studies can be 
attributed to different solar models and slightly different microturbulent
velocities (0.8 km s$^{-1}$ in Sneden \& Crocker and 1.0 km s$^{-1}$
here).  

Derived abundances depend upon the adopted 
values of effective temperature ($T_{\rm eff}$), surface gravity 
(parameterized by log $g$), overall stellar metallicity (typified by
[Fe/H]), and the microturbulent velocity ($\xi$).  In the
Smith et al. (2000) analysis, the effective 
temperatures and surface 
gravities were derived from
spectroscopy: the spectroscopic analysis used the numerous Fe I and Fe II
lines (with well-determined laboratory $gf$-values) measurable in 
high-resolution spectra.  Demanding 
the simultaneous conditions of zero slope in plots of Fe abundance
(from Fe I) versus both excitation potential and equivalent width
yield the effective temperature and microturbulent velocity.
Enforcing ionization
equilibrium, such that both Fe I and Fe II yield the same abundance,
provides the model surface gravity.  Norris \& da Costa (1995) used a
different approach to derive stellar parameters, with the 
effective temperatures determined from the infrared 
photometry of Persson et al. (1980) and Paltoglou \& Norris (1989),
using the T$_{\rm eff}$--IR color scales from Cohen, Frogel, \&
Persson (1978).  

A comparison of the derived stellar parameters in the two independent
studies (Norris \& Da Costa 1995 and Smith et al. 2000) is useful and 
is possible for 4 stars in common to both samples:
ROA's 102, 213, 253, and 219.  Mean differences and standard deviations
of the basic stellar parameters, in the sense of `Smith et al. minus
Norris \& Da Costa', were computed and these
are found to be: $\Delta$T$_{\rm eff}$= +25$\pm$125K, 
$\Delta$(log g)= -0.02$\pm$0.25 dex, 
$\Delta$$\xi$= +0.10$\pm$0.27 km s$^{-1}$, and
$\Delta$[Fe/H]= -0.05$\pm$0.06 dex.  Therefore, there are no significant 
offsets between the stellar parameters derived by Norris \& Da Costa 
(1995) and those from Smith et al. (2000), with the scatter, as measured 
by the standard deviations, being very close to the estimated uncertainties
as discussed in both papers.  

An additional check of the Smith et al. (2000) Fe I--T$_{\rm eff}$ scale
against IR color scales is possible using data from the Two Micron
All Sky Survey (2MASS: http://www.ipac.caltech.edu/2mass/).
Infrared J, H, and K magnitudes can be found for 8 of the
Smith et al. stars and effective temperatures were computed for these
giants using four different (V--K)--T$_{\rm eff}$ calibrations, from
Cohen et al. (1978), McWilliam (1990), van Belle et al. (1999), and
Alonso, Arribas, \& Martinez-Roger (1999).  Transformations of the 2MASS
colors, to other photometric systems, can be found in
Carpenter (2001) and, for the $\omega$ Cen stars, are quite small.
The four different (V--K)
calibrations yield effective temperatures with scatters of about $\pm$60K
for an inidividual $\omega$ Cen giant, and the comparison of the mean
difference and standard deviation of T$_{\rm eff}$(Fe I) -- 
T$_{\rm eff}$(V--K) is +38$\pm$138K: this is very similar to the
comparison to the Norris \& Da Costa (1995) T$_{\rm eff}$'s. 
Due to the lack of any
significant offsets, we expect no systematic differences as the previously
derived stellar parameters from Norris \& Da Costa and Smith et al.
are on a consistent scale within the errors.
Therefore, the adopted parameters and Fe abundances 
for the $\omega$ Cen sample stars remain the same from
the previous studies and these are assembled in Table 3
together with the derived copper abundances for the sample stars
studied in this cluster. 

In addition, the other sample stars from the globular clusters 
NGC's 288, 362, 3201,
6752, and M4 are analyzed such that their stellar parameters are 
derived in the same way as by
Smith et al. (2000) and therefore rely on the spectral lines of Fe I 
and Fe II present in their spectra.
Tables 4
and 5 list the Fe I and Fe II measured equivalent widths for 
these 15 giants (as they have not been published previously). 
Their corresponding 
stellar parameters and resulting Fe and Cu abundances are
presented in Table 6.
 
We note that  some of the stars from our globular clusters sample
have been examined previously (using the same spectra)
by Ivans et al. (1999), for M4, and Gonzalez \& Wallerstein (1998),
for NGC3201 (with neither study including copper in their analyses).  
These authors used either a different technique (Ivans et al. used
line-depth ratios), or a somewhat different set of Fe I and Fe II lines 
(Gonzalez \& Wallerstein 1998) to derive parameters.  A comparison of their
adopted parameters is useful as an indication of the magnitude of the 
uncertainties involved in different spectroscopic determinations. For
the four M4 giants, differences of (This Study - Ivans et al.) results in
$\Delta$T$_{\rm eff}$= +105$\pm$43K,
$\Delta$(log g)= -0.06$\pm$0.16 dex,
$\Delta$$\xi$= +0.02$\pm$0.10 km s$^{-1}$, and
$\Delta$([Fe/H])= +0.04$\pm$0.07.
The comparison between (This Study - Gonzalez \& Wallerstein) finds (with
only average differences listed, as there are only 2 stars in the
comparison)
$\Delta$T$_{\rm eff}$= +125K,
$\Delta$(log g)= +0.25 dex,
$\Delta$$\xi$= +0.10 km s$^{-1}$, and
$\Delta$([Fe/H])= -0.08 dex.

In general terms, these various comparisons between independent studies
indicate that, for these red giants, 
effective temperatures can be defined to within $\sim$100K, surface gravities
to about 0.1-0.2 dex, microturbulent velocities within 0.1-0.2 km s$^{-1}$,
and metallicities to within about 0.05 dex.  In terms of copper abundances,
typical errors of 100K in T$_{\rm eff}$, 0.2 dex in log g, and 
0.2 km s$^{-1}$ in $\xi$ would lead to expected uncertainties of 0.18 dex
in derived Cu abundances.

Sample spectral fits of the Cu I line regions are
shown in Figures 2, 3, and 4.  The star pe75 from NGC362 is shown in
Figure 2, illustrating one of the globular cluster giants (other than
$\omega$ Cen) observed from CTIO.  Note that the Cu I
line is blended with a Cr I line, although the Cu I line itself is
well-defined.  Synthetic spectra are plotted for three different Cu
abundances: the best-fit value, plus-or-minus 0.1 dex values.  Figure 3
also shows a CTIO spectrum, but for the $\omega$ Cen star ROA 324 (a star
with a rather low [Cu/Fe] ratio).   Finally, Figure 4 shows an AAT
spectrum and resulting synthetic fits for the $\omega$ Cen giant
ROA 252 (with [Fe/H]= -1.74). 

Because some of the $\omega$ Cen giants are fairly cool (T$_{\rm eff}$
as low as 3750K), a check on the model atmospheres was performed
using the latest version of the MARCS code, OSMARCS (Plez, Brett, \&
Nordlund 1992; Edvardsson et al. 1993; Asplund et al. 1997), that
contains more extensive and up-to-date molecular opacities.  An OSMARCS
model was computed for the coolest giant, ROA201, and the Cu I region
was synthesized and analyzed with the newer model as a comparison to the
older MARCS model.  Virtually no differences were found in the respective
synthetic spectra and the same copper abundance was found from both models.
This exercise demonstrates that over the temperature range of giants 
studied here the Cu abundance scale is insensitive to the older MARCS
versus newer OSMARCS atmospheres.

As a final check on the copper and iron abundance presented here, possible
systematic effects, especially with temperature, were investigated.  
In low-mass red-giant branches (RGB's),
the position of the giant branch in a T$_{\rm eff}$--luminosity plane 
is a function of metallicity, with the more metal-rich red giants
being cooler.  Since both the Norris \& Da Costa (1995) and Smith
et al. (2000) samples were observed, essentially, at near-constant 
V-magnitudes while crossing
the RGB in color, the more metal-rich stars tend to be
cooler in both samples.  Such a correlation could lead to systematic
effects in [Cu/Fe] versus [Fe/H].  Possible systematic trends
correlated with stellar parameters are investigated via Figure 5.  In
the top panel is a spectroscopically derived HR-Diagram, with log g
versus T$_{\rm eff}$: it is clear that the lower surface gravity stars
tend to be cooler, as would be expected for low-mass stars ascending the
RGB.  Superimposed on the observationally derived points are two sample
isochrones from Bergbusch \& Vandenberg (1992) for 12 Gyr populations
with [Fe/H]= -2.0 and -1.0, respectively.  These iron abundances should
nearly bracket the $\omega$ Cen stars and, indeed, this is the case.
The close match between the derived stellar parameters and those
from stellar models suggests that there are not large systematic offsets
in the temperatures and gravities used to derive the abundances.

The middle panel of Figure 5 shows [Fe/H] versus T$_{\rm eff}$ and
demonstrates that the coolest red giants in both samples tend to be
found among the most metal-rich stars (with these [Fe/H] values being
those derived spectroscopically): this is the expected result.  Finally,
the bottom panel shows [Cu/Fe] versus effective temperature and no
significant trend exists.  The straight line is a least-squares
fit to the points and has but an insignificant non-zero slope; there is
no evidence that there are measurable systematic abundance trends created
by the derivation of stellar parameters or the abundance analysis.  
It should be noted that if $\omega$ Cen stars followed the behavior of
the halo one would expect a trend, i.e. the more metal rich (and cooler)
giants would have larger values of [Cu/Fe]. It would be unlikely that a
systematic temperature error would just conspire to provide the observed zero
slope of [Cu/Fe] with effective tempetarure. It is more likely that
the stellar parameters are not affected by significant systematic errors
and that
the [Cu/Fe] ratio in $\omega$ Cen is quite constant from [Fe/H]=-2.0
to -0.8. A combination of all of the [Cu/Fe] values for the
$\omega$ Cen stars results in a mean and standard deviation of
-0.50 $\pm$ 0.11: this indicates no evolution in the copper to iron
yields over much of the chemical evolution within $\omega$ Cen.

\section{Discussion}

The abundances of copper discussed in this paper are assembled in the
different panels of Figure 6 as [Cu/Fe]
versus [Fe/H].  As a reference point, we adopt log $\epsilon$(Fe)= 7.50
and log $\epsilon$(Cu)= 4.06 as the corresponding solar abundances.  
All three samples of stars are shown in
Figure 6: the field star results, the
``mono-metallicity'' globular clusters (NGC's 288, 362, 3201, 6752, and
M4), and $\omega$ Cen.
 
In the top panel of Figure 6 are shown the results from Sneden \&
Crocker (1988) and Sneden et al. (1991) 
for the field stars, with their corresponding error estimates in [Cu/Fe]
and [Fe/H] plotted.  The straight lines are least-squares fits to two
metallicity regimes ([Fe/H]$\le$ -1 and [Fe/H]$\ge$ -1) and are
meant only to serve as guides to the eye
in order to more easily identify the behavior of copper with metallicity
([Fe/H]). 
It is found that [Cu/Fe] is effectively constant in the high-metallicity 
regime while clearly decreasing towards lower metallicities in those
stars having [Fe/H]$\le$ -1.4
(as discussed in Sneden et al. 1991), with an
approximate slope of 0.4 in [Cu/Fe] versus [Fe/H].  

The middle panel of Figure 6 again shows the field-star results (with
error bars suppressed for clarity), upon which are plotted our abundances
for the globular clusters M4, NGC's 288, 362, 3201, and 6752.  
Note that the individual scatter, in both [Fe/H] and
[Cu/Fe], within each globular cluster is quite small; thus, for example,
the small differences in [Cu/Fe] between NGC362 and the clusters M4 and
NGC288, at similar values of [Fe/H], are probably
real, with the three studied stars in NGC 362 falling slightly below
the straight line that represents the field behavior. (Although there
is a gap in the sample field population at roughly this metallicity).  
Except for NGC362, that may have a somewhat low [Cu/Fe] ratio
for its value of [Fe/H], the other 4 globular clusters fall quite close
to the mean trend of the field stars.  

The general trend for the $\omega$ Cen stars is shown in the bottom
panel of Figure 6, again with the field-star results overplotted
for comparison.  Unlike the field, the $\omega$ Cen members show 
a constant [Cu/Fe] ratio of about -0.5 in the interval from
[Fe/H] of $\sim$ -2.0 up to -0.8: there may be a hint that the scatter in 
[Cu/Fe] increases towards the highest metallicities. The values of [Cu/Fe]
in the field halo stars and $\omega$ Cen are in agreement at [Fe/H]$\sim$ -2, 
where the two
samples overlap, but are clearly different at 
[Fe/H]$\sim$ -1, with the $\omega$ Cen stars falling below the field
in [Cu/Fe].  
In particular, over the upper range of metallicities of the $\omega$ Cen
sample studied here (with -1.2 $\le$ [Fe/H] $\le$ -0.8) the mean value
of [Cu/Fe] and standard deviation
is -0.52 $\pm$0.17 (for 9 stars). For the same [Fe/H] range in the
field-star sample, [Cu/Fe]=-0.05 $\pm$0.18 (for 8 stars).
The nearly horizontal line in this panel is a least-squares
fit to the $\omega$ Cen points that has a near-zero slope.  
The 8-pointed stars are the results for $\omega$
Cen from Pancino et al. (2002), and include two members that are
significantly more metal-rich than those in our sample.  For the stars
with [Fe/H]$\le$ -0.8, the Pancino et al. (2002) results overlap those 
found in this study; however, $\it{one}$ of the metal-rich Pancino et al. 
(2002) stars
(with [Fe/H]$\sim$ -0.5) has a [Cu/Fe] value that is nearly equal to
those found in the field stars at this metallicity. 
A more rigorous comparison between $\omega$ Cen and field behavior
will be aided by an increased sample of galactic field stars
allowing for a complete elucidation of the behavior of copper in
the metallicity range between [Fe/H] -2.0 and -0.5.

From the point of view of observations, the behavior of [Cu/Fe] is 
different between the field halo stars and the stars in $\omega$ Cen
over the studied metallicity range. 
Can this different behavior in the
two samples be understood within some simple context?  It is safe
to assume that the production site(s) for Cu involve either massive
stars and Type II supernovae, or Type Ia supernovae (which result
from binary-star evolution), and do not involve nucleosynthesis
in low-mass AGB stars.  Within this context, $\omega$ Cen shows 
clear evidence of enormous
elemental contributions from low-mass AGB stars via s-process
nucleosynthesis (e.g. Norris \& Da Costa 1995), thus, the observed
constant values of [Cu/Fe] found here do not fit a picture in
which Cu synthesis arises in AGB stars.  Addressing the point of
whether Cu production is associated with either SN II or SN Ia,
there are conflicting ideas.  As mentioned in the Introduction,
based on their observed trend of
[Cu/Fe] $\alpha$ 0.4[Fe/H], Sneden et al. (1991) favor a production
model for Cu in metal-poor stars ([Fe/H] $\le$ -1.0) which has two
sources, both occurring in massive stars: the weak s-process, operating
during core He burning, plus the e-process of Si burning during
Type II supernovae.  In the more metal-rich disk stars, an additional
source of Cu appears from Type Ia supernovae.  On the other hand,
Matteucci et al. (1993)
use the Cu--Fe relation from Sneden et al. (1991) to reach
somewhat different conclusions about the origin of Cu.  Their chemical
models are best fit with Cu being produced predominantly in Type Ia
supernovae, even down to metallicities well below -1.0 in [Fe/H].
One of the main points of their study is to assume that there is
already a SN Ia contribution between metallicities [Fe/H] -2.0 and -1.0.

A number of previous studies of $\omega$ Cen (e.g. Brown \&
Wallerstein 1993; Norris \& Da Costa 1995; Smith et al. 1995, 2000)
find that elements such as Si, Ca, or Ti exhibit the same behavior
as observed in the halo field, i.e. that their abundances are
enhanced relative to Fe.  This is interpreted to result from
chemical enrichment from Type II supernovae.  Since the halo field
and $\omega$ Cen are identical in this respect, there is evidence
that $\omega$ Cen was enriched by SN II ejecta.  Thus, the fact
that [Cu/Fe] exhibits different behavior between the field-halo stars
and $\omega$ Cen could suggest that the main source of Cu within this
metallicity range is not massive stars.  Of course, one could
speculate that, for example, the weak s-process operates over some
restricted mass range and that the mass function of $\omega$ Cen
was deficient in stars over this range.  This speculation, however,
introduces additional parameters and will not be explored further
here.

If the [Cu/Fe] ratios in $\omega$ Cen do indeed eliminate massive stars as the
dominant Cu producers, this leaves Type Ia supernovae, as suggested
by Matteucci et al. (1993).  This winnowing of Cu sources still
leaves the problem of why the field halo and $\omega$ Cen exhibit 
different behaviors of [Cu/Fe] versus [Fe/H], but there are a number
of possible explanations: 1) The timescale for SN Ia progenitor
evolution exceeds the timescale of chemical evolution within $\omega$
Cen, 2) SN Ia systems do not form in metal-poor environments, as
suggested by Kobayahsi et al. (1998), 3) SN Ia progenitor systems
(presumed to be binaries) are sufficiently rare, that none would be
expected to form in a small system, such as $\omega$ Cen, 4) SN Ia binary
progenitor systems form, but are subsequently disrupted 
in the dense environment of a globular
cluster, or, 5) the negligible contribution from SN Ia's over much of
$\omega$ Cen's chemical evolution may result from the selective retention 
of stellar ejecta in low-mass systems (where AGB winds are more efficiently
retained than SN II ejecta, and, in the case here, SN Ia ejecta almost
not retained at all). 

Concerning option (1), 
photometric studies by Hilker \& Richtler (2000) and Hughes \& 
Wallerstein (2000), as well as the s-process spectroscopic analyses 
by Smith et al. (2000), indicate that it took $\sim$ 1-3 Gyr for $\omega$
Cen to evolve from [Fe/H]$\sim$ -2.0 to -1.0.
The nucleosynthesis from Type II supernovae and from the weak
component of the s-process will happen on time scales of 10$^{7}$ years.
The timescale for the formation of Type Ia supernovae is not clear
mostly due to the lack of a definite progenitor model. The
discussion by Kobayashi et al. (1998) gives typical time scales for the
double degenerate model as 0.3 Gyr and for single degenerate models of
0.6 Gyrs. A discussion of a larger set of possible Type Ia models gives
rough timescales from 0.1 to 4 Gyrs (Branch et al. 1995). It has been
long established that the rates of Type Ia supernovae are higher in
late-type galaxies (Oemler \& Tinsley 1979) than in early types,
which would support a shorter time scale for formation of Type Ia
events. It therefore seems safe to assume that the age spread measured
in the stellar population $\omega$ Cen is significantly longer than the time
needed to form the typical Type Ia supernovae.

Concerning option (2), 
Kobayashi et al. (1998) develop models for binary SN Ia progenitors in
which the wind from the white dwarf plays a key role in determining
whether a Chandrasekhar mass limit is reached. The strength of the wind
is a function of the metallicity, and Kobayashi et al. (1998) conclude
that Type Ia supernovae are quenched below metallicities of -1.0.
In this type of picture, it might be expected that SN Ia's would be
inhibited in all low metallicity environments and thus the field halo and 
$\omega$ Cen would exhibit the same type of behavior. However, this is
not what is observed. But of course, this contention resides on the 
assumption that
Ia's are the main producers of Cu.

Concerning possibility (3) from above, Pagel (1997) estimates that the
SN Ia rate in the Milky Way is 0.33 per century.  Taking a Galactic
stellar mass of
$\sim$ 10$^{11}$M$_{\odot}$, this translates to a ``mass specific rate''
of 3 x 10$^{-14}$ SN Ia events per year per solar mass of stars.  Given
an $\omega$ Cen protomass of $\sim$10$^{7}$M$_{\odot}$, it can be estimated
that the SN Ia rate in the cluster would be 10$^{-7}$ events per year
once the time has passed for SN Ia events to turn on.
Over some 10$^{9}$ years of chemical evolution, $\omega$ Cen would
have $\sim$ 100 SN Ia's occur.  As each SN Ia can eject up to
$\sim$0.6M$_{\odot}$ of Fe (Pagel 1997), these events would have 
a measurable impact
on the abundances, if such ejecta were retained.  Of course, this
simple argument assumes that the rate of binary formation in a globular
cluster is the same as in the Galaxy as a whole.  There is no observational
evidence concerning the initial fraction of binary systems formed in
globular clusters; however, there is evidence that binary sytems can
be disrupted in the environment of a globular cluster.  
In fact,
C\^ot\'e et al. (1996) find that $\omega$ Cen has a smaller
fraction of binary systems than other globular clusters studied by
them (e.g. M71, M4 and NGC3201 among others). This could be a possible  
explanation for the lack of SN Ia enrichment over much of the [Fe/H]
evolution in $\omega$ Cen, as suggested by possibility (4) from above.
In this context, perhaps the incidence of Type Ia supernovae is
simply lower in $\omega$ Cen compared to the field due to some
destruction of SN Ia progenitor systems. This might explain the
delayed, but eventual appearence of SN Ia enrichment in the most metal-rich
stars in $\omega$ Cen ([Fe/H] $\sim$ -0.5) as derived by Pancino
et al. (2002).

Another possibility (number 5 from above) in understanding these 
peculiar chemical
traits in $\omega$ Cen comes from Smith et al. (2000).
The observation that $\omega$ Cen contains such a large
s-process component led to the speculation that in
a relatively low-mass stellar system AGB ejecta, because of their low
velocity winds, might be  more
efficiently retained relative to the much faster moving
Type II supernova ejecta.  The AGB ejecta would then be mixed 
with a relatively small amount of SN II ejecta that was retained, 
as well as with the residual gas of initial composition.
The same sort of speculative explanation could hold for explosions
of Type Ia supernovae. Differing retention efficiencies between
SN II and SN Ia ejecta might be caused by the differing interstellar
environments where the explosions occur.
For example, Recchi, Matteucci \& D'Ercole (2001) model chemical
and dynamical evolution in a low-mass, starburst dwarf galaxy
and find that a metal-enriched  galactic wind is driven out of
the galaxy by SN II and SN Ia explosions. They also find that the
SN Ia ejecta are preferentially ejected relative to SN II ejecta.
Gnedin et al. (2002) have pointed out, however,
that the current escape velocity from $\omega$ Cen is
not that different from the other globular clusters. The hypothesis of
selective mass loss or retention, on the other hand, applies to the progenitor
object (proto $\omega$ Cen) before it was captured into its current 
retrograde orbit around the Milky Way;
the nature of that progenitor object remains unknown.
Distinguishing whether possibility (4) or (5) best pertains to 
understanding
the chemical evolution within $\omega$ Cen can probably be
constrained by determining abundance ratios in a wide variety of
stellar environments, such as other types of small galaxies within
the Local Group, as well as in a larger sample of $\omega$ Cen
members from the most metal-rich population identified by Pancino
et al. (2000).  

Our abundance results lead us to the conclusion that the copper to iron ratio
in $\omega$ Cen is nearly constant over about a factor of 10-15 range in
iron abundance (-2.0 $\le$ [Fe/H] $\le$ -0.8). Over this same range  
of [Fe/H] in the field stars, [Cu/Fe] increases by about 0.5 dex. 
This demonstrates yet another peculiar
trait of the chemical evolution found within $\omega$ Cen when compared
to the Galactic halo field stars at comparable metallicities (the other
peculiar chemical trait being the large s-process abundance component
in $\omega$ Cen). 

\section{Conclusions}

In this work we discuss Cu abundances in 40 red-giant members of
$\omega$ Cen, as well as 15 red-giant members of 5 other globular clusters
that overlap the range of [Fe/H] found in $\omega$ Cen.  Abundances
from field stars already published in the literature by Sneden \&
Crocker (1988) and Sneden et al. (1991) are included in the discussion.
It is found that both the field star sample and 4 of the 5 
``mono-metallicity''
Galactic globular clusters display similar values of [Cu/Fe]
at a given [Fe/H].  The $\omega$ Cen members, however, tend to
have smaller and nearly-constant values of [Cu/Fe] ratios, when compared to the other globulars
and the field, as [Fe/H] increases.  If copper is produced preferentially
in SN Ia, relative to SN II, this indicates that $\omega$ Cen stars show
little, if any, enrichment from SN Ia's up to a cluster metallicity of
[Fe/H]$\sim$ -0.8.  
A lag in SN Ia chemical enrichment in $\omega$ Cen could result either
from the disruption of binary SN Ia progenitor systems in $\omega$ Cen,
or from a higher fraction of SN Ia ejecta being lost from $\omega$ Cen.

This work is supported in part by the National Science Foundation through
AST99-87374 (VVS).

\clearpage

\figcaption[fig1.ps]{Spectral syntheses of the observed Cu I line 
in the solar
flux atlas (Kurucz et al. 1984) for different copper abundances.
The synthetic Cu I line-profile produced from the hyperfine and isotopic
components in Table 2 produces a good fit to the observed solar profile
for log $\epsilon$(Cu)=4.06. The synthetic profiles are convolved 
with broadening from the instrumental profile (as discussed by
Kurucz et al. 1984), solar rotation, and required additional broadening,
assumed to consist of gaussian distributed macroturbulence ($\Gamma$).
\label{fig1}}

\figcaption[fig2.ps]{The Cu I spectral region for a red giant in the 
globular cluster NGC 362.  This illustrates one of the CTIO spectra and
shows three different copper abundances (each differing by 0.10 dex) in
the synthetic spectra.
\label{fig2}}

\figcaption[fig3.ps]{The Cu I line in an $\omega$ Cen red giant: this
target is one of the most metal-rich stars in this sample ([Fe/H]= -0.95) 
and is one of the CTIO specta.  Three different copper abundances are shown
in the synthetic spectra, illustrating the sensitivity of the Cu I
line-strength to the copper abundance.
\label{fig3}}

\figcaption[fig4.ps]{An example of an AAT spectrum and the synthetic
fits.  This is a fairly metal-poor red giant in $\omega$ Cen,
with [Fe/H]= -1.74.  The typical AAT spectra have signal-to-noise 
ratios that are somewhat
less than in the CTIO spectra (as can be seen by comparing this spectrum
to those in Figures 2 and 3), but more stars are contained in the AAT
dataset.
\label{fig4}}

\figcaption[fig5.ps]{A search for possible trends in combinations of the
derived stellar parameters and abundances.  The top panel is a
version of an HR-diagram in which surface gravity is plotted versus
effective temperature; as red giants of nearly the same mass ascend
the giant branch towards lower values of T$_{\rm eff}$, their surface
gravities should decrease (as observed).  The solid curves are sample
isochrones (from Bergbusch \& Vandenberg 1992) for 12 Gyr populations
having two different metallicities.  The middle panel shows derived
Fe abundances versus T$_{\rm eff}$: note the trend of lower effective
temperatures as [Fe/H], which is expected for low-mass red giants (the
red-giant branch shifts towards cooler temperatures as metallicity
increases).  The bottom panel plots [Cu/Fe] against T$_{\rm eff}$.
This distribution is essentially flat, as illustrated by the fitted
least-squares line which has a near-zero slope (there is slight,
non-significant slope).  There are no trends in the derived copper
iron ratios with temperature.
\label{fig5}}

\figcaption[fig6.ps]{The ratios of [Cu/Fe] versus [Fe/H] for three samples
of stars:  the field stars from Sneden \& Crocker (1988) and 
Sneden et al. (1991) are shown as open circles in all three panels, 
while members of the mono-metallicity
globular clusters NGC's 288, 362, 3201, and 6752, and M4 are plotted
in the middle panel and the $\omega$ Cen giants are shown in the bottom 
panel.
The straight lines through the field stars are linear least-squares fits
to two separate regimes ([Fe/H] $\le$ -1.1 and $\ge$ -1.1).  The field-star
results show no significant evolution in [Cu/Fe] versus [Fe/H] in the
high-metallicity sample, but show decreasing [Cu/Fe] with decreasing
[Fe/H] in the low-metallicity sample (with a slope in this type of plot
of 0.4).  The mono-metallicity globular cluster stars shown in the
middle panel generally fall well within the scatter in [Cu/Fe] exhibited
by the field stars.  The cluster NGC 362 may be slightly different,
falling slightly below the field-star distribution: this cluster deserves
future attention.  The copper to iron ratios for $\omega$ Cen are shown 
in the bottom panel, and exhibit a very flat distribution of [Cu/Fe]
over the metallicity range covered by the sample of stars analyzed here
(with [Fe/H]= -2.0 up to -0.8).  This behavior is different from the 
field stars over this metallicity range, which show increasing [Cu/Fe]
over this same range in [Fe/H].  The more metal-rich giants in 
$\omega$ Cen, analyzed recently by Pancino et al. (2002), also tend to
follow this flat distribution of [Cu/Fe], except for the two most
metal-rich members, which show increasing copper to iron ratios. 
\label{fig6}}
\end{document}